\documentclass[journal]{IEEEtran}
\usepackage{amsmath,amssymb,amsfonts}
\usepackage{graphicx}
\usepackage{cite}
\usepackage{stfloats}
\usepackage{wrapfig}
\usepackage{floatrow}
\usepackage{textcomp}
\usepackage{xcolor}
\usepackage{algorithm}
\usepackage{algpseudocode}

\def\BibTeX{{\rm B\kern-.05em{\sc i\kern-.025em b}\kern-.08em
    T\kern-.1667em\lower.7ex\hbox{E}\kern-.125emX}}

\begin{document}

\title{QUEST: A Quantized Energy-Aware SNN Training Framework for Multi-State Neuromorphic Devices}
\author{
    \IEEEauthorblockN{
        Sai Li, \textit{Member, IEEE},
        Linliang Chen,  
        Yihao Zhang,
        Zhongkui Zhang,
        Ao Du, \textit{Student Member, IEEE}\\ 
        Biao Pan, 
        Zhaohao Wang,
        Lianggong Wen, \textit{Member, IEEE}}
        and Weisheng Zhao, \textit{Fellow, IEEE}

\thanks{Manuscript received XX XX, 2025; revised XX XX, 2025. This work was supported by the National Natural Science Foundation of China under Grant 62404015, the Fundamental Research Funds for the Central Universities and the Beijing Outstanding Young Scientist Program

\textit{(}Sai Li and Linliang Chen contribute equally to this work.\textit{) (}Corresponding author: Lianggong Wen, Weisheng Zhao\textit{)}}
\thanks{S. Li is with the School of Integrated Circuit Science and Engineering, and also with the School of Computer Science and Technology, Beihang University, Beijing 100191, China}
\thanks{L. Chen, Z. Zhang, Y. Zhang, A. Du, B. Pan, Z. Wang, L. Wen, and W. Zhao are with the School of Integrated Circuit Science and Engineering, Beihang University, Beijing 100191, China}
}

\maketitle

\begin{abstract}
Neuromorphic devices, leveraging novel physical phenomena, offer a promising path toward energy-efficient hardware beyond CMOS technology by emulating brain-inspired computation. However, their progress is often limited to proof-of-concept studies due to the lack of flexible spiking neural network (SNN) algorithm frameworks tailored to device-specific characteristics, posing a significant challenge to scalability and practical deployment. To address this, we propose QUEST, a unified co-design framework that directly trains SNN for emerging devices featuring multilevel resistances. With Skyrmionic Magnetic Tunnel Junction (Sk-MTJ) as a case study, experimental results on the CIFAR-10 dataset demonstrate the framework’s ability to enable scalable on-device SNN training with minimal energy consumption during both feedforward and backpropagation. By introducing device mapping pattern and activation operation sparsity, QUEST achieves effective trade-offs among high accuracy (89.6\%), low bit precision (2-bit), and energy efficiency ($\sim$\text{93}$\times$ improvement over the ANNs).  QUEST offers practical design guidelines for both the device and algorithm communities, providing insights to build energy-efficient and large-scale neuromorphic systems.
\end{abstract}

\begin{IEEEkeywords}
Spiking Neural Network (SNN), Neuromorphic Computing, Neuromorphic Device, Quantization, Non-Volatile Memory, On-device Training
\end{IEEEkeywords}

\section{Introduction}
\IEEEPARstart{S}{piking} Neural Networks (SNNs) represent a paradigm shift in neural computing, offering distinct advantages over conventional artificial neural networks (ANNs) \cite{eshraghian2023training,hassabis2017neuroscience,maass1997networks}. Using event-driven processing and temporal information encoding through discrete spikes, SNNs achieve remarkable energy efficiency while closely mimicking biological neural systems\cite{frenkel2023bottom,rathi2021diet,zhang2021tuning}. This biological plausibility, combined with their sparse and binary communication mechanism, makes SNNs particularly compelling for edge computing and real-time applications where resource constraints are critical\cite{zhang2020fully,kim2023c}.

In recent years, the emergence of neuromorphic chips has further accelerated the development of SNNs by providing efficient implementation platforms \cite{pei2019towards,davies2018loihi,merolla2014million}. Specifically, various emerging nonvolatile memory (NVM) devices—including Magnetoresistive Random-Access Memory\cite{grollier2020neuromorphic}, Resistive Random-Access Memory\cite{zhang2018sign,lele2023heterogeneous,yang2024fully}, Phase Change Memory\cite{wang2021phase,jiao2020monatomic}, Electrochemical Random Access Memory\cite{tang2018ecram}, and Ferroelectric Field-Effect Transistor\cite{luo2019capacitor}—as well as novel material platforms such as moiré heterostructures\cite{yan2023moire}, which have demonstrated remarkable potential in implementing key components of SNNs. These neuromorphic hardware platforms typically support basic neuron models like Leaky Integrate-and-Fire (LIF) and Integrate-and-Fire (IF), and with more sophisticated implementations incorporate biological features such as hard/soft reset mechanisms, adaptive threshold\cite{zhang2023low} and refractory periods\cite{burkitt2006review,massarotto2024adiabatic}. At the synaptic level, multi-state resistive devices can effectively emulate synaptic weights, with some devices exhibiting bio-inspired learning mechanisms like Spike-Timing-Dependent Plasticity (STDP)\cite{yao2023tunneling} and advanced neoHebbian synaptic models\cite{wang2020neuromorphic,bhattacharya2023reram}.

Existing tools and frameworks have made significant strides in enabling the co-development of neuromorphic algorithms and hardware, though their areas of focus have varied. For example, NeuroSim (versions 1.0 and 2.0) primarily targets the co-design of ANNs with multi-level resistive devices, emphasizing compute-in-memory inference and on-chip training\cite{peng2020dnn+, chen2017neurosim+}; however, it does not support SNNs. Software frameworks such as Norse\cite{pehle2021norse}, snnTorch\cite{eshraghian2023training}, and SpikingJelly\cite{fang2023spikingjelly} focus on developing SNN algorithm modules and supporting mature CMOS-based neuromorphic chips, but they lack support for emerging NVM devices. Abisko provides architectural-level abstractions\cite{vetter2023abisko} without direct code or device integration.

\begin{figure}[t]
\includegraphics[width=1\columnwidth]{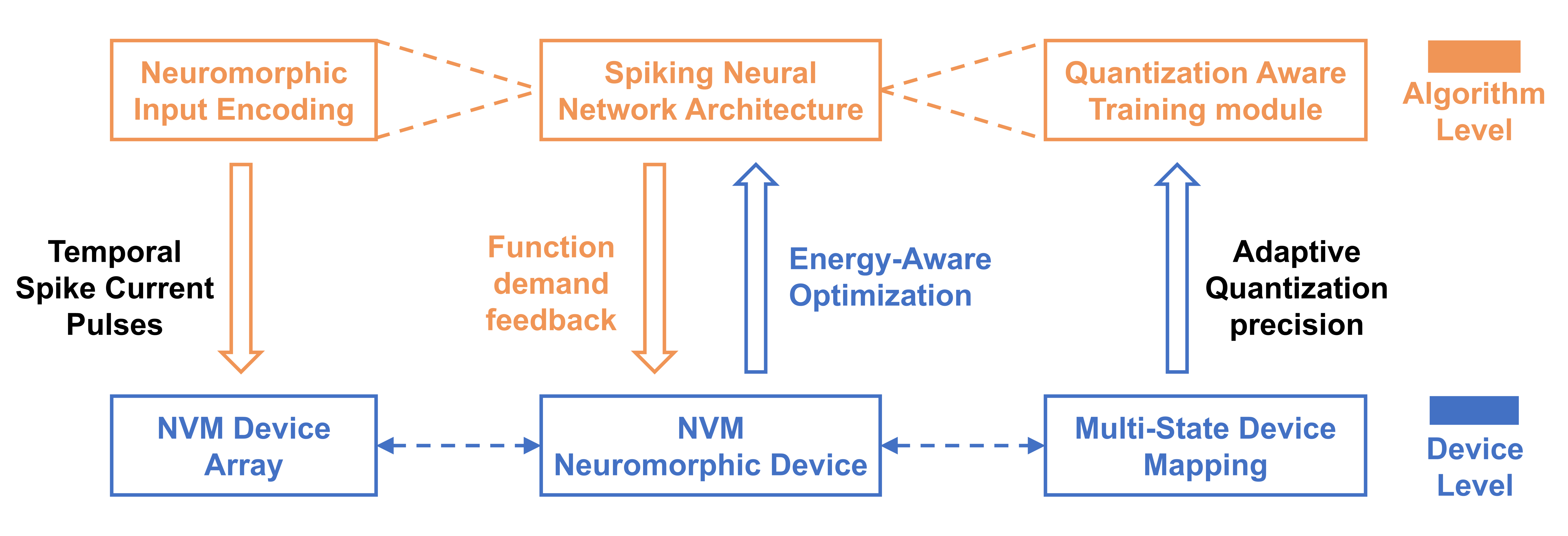}
\caption{Hierarchical architecture of the QUEST framework illustrating the interplay between algorithm-level (orange) and device-level (blue) implementations. The framework enables efficient co-optimization through bidirectional interactions, including encoding, function feedback, energy-aware optimization, and adaptive quantization mechanisms.}
\label{Fig1}
\end{figure}

Despite these efforts, a significant challenge remains: the disconnect between device research and algorithm development. This divide, resulting from differing optimization objectives and technical approaches, impedes the realization of energy-efficient, large-scale neuromorphic systems \cite{roy2024spintronic, hamilton2021best}. The situation is further complicated by the demands of on-device training, which negatively impacts SNN time steps and sparsity, ultimately reducing energy efficiency. Three key issues are often neglected:

(1) Mismatch between SNN encoding schemes and device input characteristics: While many encoding schemes are algorithmically efficient, they may lead to an increased number of time steps when implemented on-chip. To ensure optimal performance, the specific input requirements and limitations of the target hardware must be carefully considered during the design of encoding methods.

(2) Influence of weight update dynamics on energy efficiency: The impact of frozen versus dynamic weights during SNN training is frequently overlooked. It can result in suboptimal energy utilization, as frozen weights do not require write operations for updates in NVM neuromorphic devices, whereas dynamic weights will cause delay and energy consumption\cite{zhang2024hyb}.

(3) Uncharted territory: The complex interplay between post-quantization energy efficiency, weight sparsity, and spike rate—particularly when scaling SNNs beyond simple datasets like MNIST—remains largely unexplored. A co-design approach that simultaneously optimizes these interconnected factors is crucial for maximizing energy efficiency in more complex applications.

To bridge this gap, we present QUEST, an open-source co-design framework that enables seamless integration between multi-level NVM neuromorphic devices and SNN algorithm training. QUEST implements a device-centric strategy. The key contributions of this work are as follows:

\textbf{Integration of direct encoding with device input}. We propose a training method that combines direct encoding with the device's input characteristics. This approach effectively avoids the drawbacks of other encoding schemes, such as increased device operation time and energy consumption, while providing valuable design insights for SNN encoding.

\textbf{Monitoring weight change dynamics}. We introduce a method for tracking weight changes during training, proposing a device output and weight state mapping (R-S mapping pattern). This enables the identification of the optimal device mapping, reducing energy consumption by $50\%$ during weight update.

\textbf{ Weight activation and its impact on energy efficiency}. We introduce a new concept called "weight activation," defined as Activation Operation Sparsity (AOS), to accurately describe the impact of sparsity on energy efficiency. This allows us to calculate the layer-wise energy consumption of the device during the forward propagation of a sample, guiding the design and optimization of on-device training SNN algorithms.

\section{QUEST framework principles}

Fig.~\ref{Fig1} illustrates the overarching architecture of our proposed training framework QUEST, which integrates both algorithm-level and device-level components to achieve a cohesive optimization with a given task and its corresponding dataset. QUEST begins with Neuromorphic Input Encoding, converting the dataset into temporal spike pulses using different encoding methods. These pulses are transmitted into the NVM Device Array, managed by the device's resistance mechanism. Considering that typical NVM devices do not achieve full precision (FP32), the direct training of SNNs requires an appropriate quantization strategy. Therefore, Adaptive Quantization Precision is utilized to match the quantization level with the number of resistance states of the device. The Quantization Aware Training Module works at the algorithm level, dynamically managing the quantization strategy to align hardware capabilities with the trade-off between accuracy and energy efficiency. At the device level, Multi-State Device Mapping is employed to map the NVM device to the quantized states of the neuron and synapse models. The core principle of QUEST is to leverage functional demand feedback from the algorithm level to adapt the behavior of the device for specific computational needs. Simultaneously, Energy-Aware Optimization adjusts device parameters to achieve optimal energy consumption while maintaining task accuracy.

\begin{figure*}[htbp]
\centering
\includegraphics[width=0.97\textwidth]{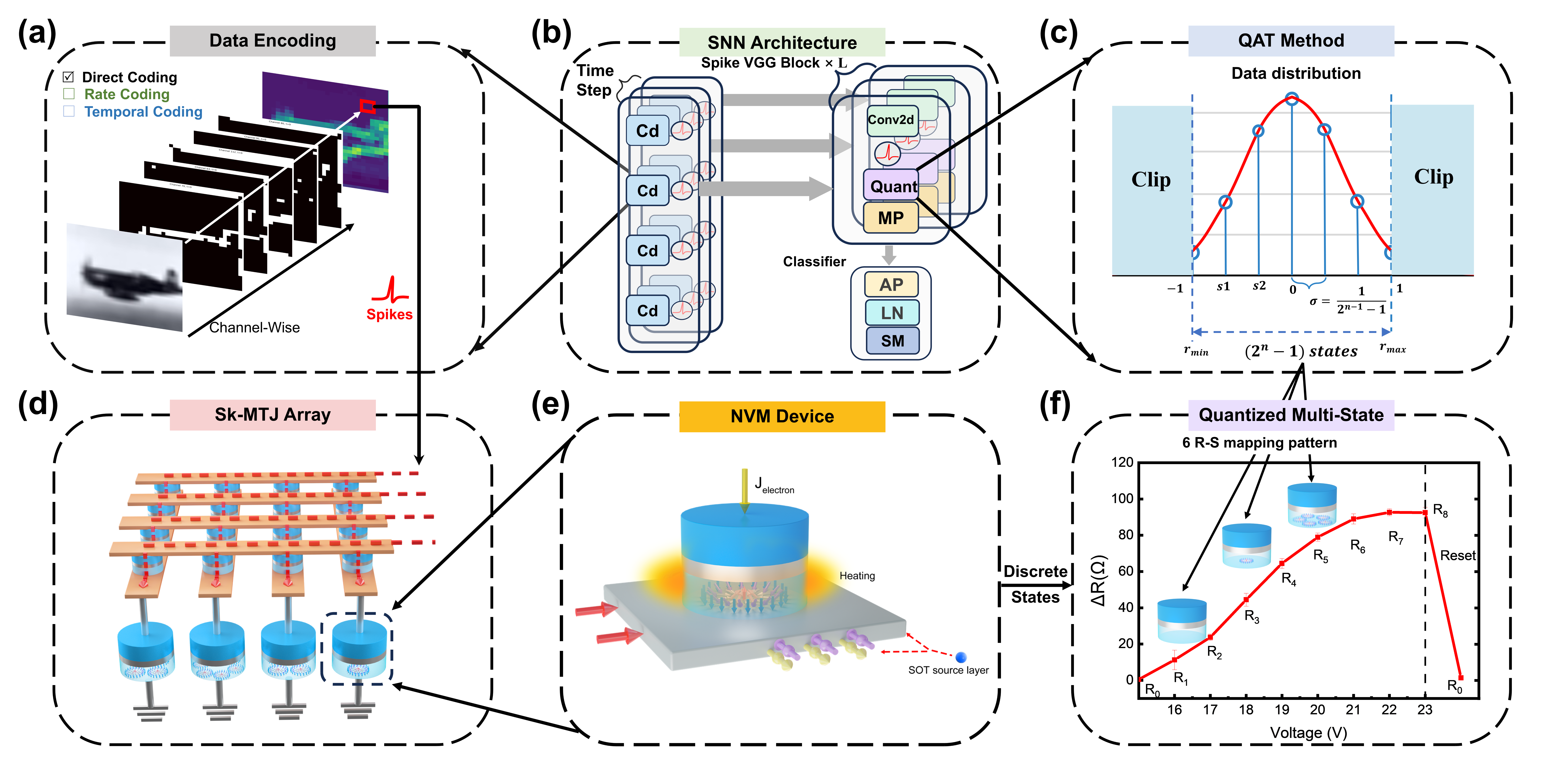}
\caption{Overview of the modules of QUEST utilized in this work. (a) Direct coding scheme for spike generation, while there are 2 other coding methods. (b) Architecture of the SNNs, featuring convolution (Conv2d), quantization (Quant), max pooling (MP), average pooling (AP), linear (LN), and softmax (SM) modules. (c) QAT method showing the data distribution with quantization step size $\sigma = 1/(2^{n-1} - 1)$, where $s_1$ and $s_2$ represent the weight state, $r_{min}$ and $r_{max}$ define the clipping range, and $(2^n - 1)$ indicates the total number of quantization states for n-bit precision. (d) Schematic of the skyrmion-MTJ (Sk-MTJ) neuromorphic array. (e) Detailed structure of a single Sk-MTJ device with write and read current pulses. (f) Resistance-voltage characteristics demonstrating the mapping between quantized states and device resistance levels.}
\label{Fig2}
\end{figure*}

Fig.~\ref{Fig2} presents an intuitive breakdown of the different modules utilized in this work. At the algorithmic level, Fig. 2b demonstrates the complete SNN training architecture, built on the spike-based VGG structure, which is suitable for image recognition tasks. Fig. 2a illustrates the data encoding process, where the direct coding method is applied to process image inputs, generating spikes at continuous time steps in a channel-wise manner. In Fig. 2c, the quantization-aware training (QAT) method is depicted, where $n-bit$ precision necessitates $2^n-1$ states to represent data, establishing the compatibility of the algorithm with hardware limitations. At the device level, Fig. 2e depicts our previously proposed NVM device-Skyrmionic Magnetic Tunnel Junction (Sk-MTJ)\cite{li2022experimental}, which can be integrated into an all-spin array to form neuron and synapse models (Fig. 2d), for processing incoming spike signals. Finally, the pulse-modulated resistance state of the Sk-MTJ device (Fig. 2f), enabling it to map directly to the quantized integer values. The mapping between resistance and integer states (R-S) follows kinds of patterns, ensuring precise alignment of the device's functionality with the algorithm's requirements. Detailed analyses of the methodological approaches for each module are presented in the following sections

\section{Device-level abstraction pipeline}
Analog neuromorphic devices have various tuning mechanisms, and there has been a general consensus that key device parameters relevant to SNN include: the number of resistance states, $\textit{R}_{on}$, on/off ratio, resistance increase/decrease pulses, resistance variation. However, during our investigation into training SNNs, we found that the mapping between device resistance states and the corresponding weights has often been overlooked or not sufficiently addressed. To address this gap, we propose a novel matrix-based mapping method that translates the resistance states of the device into the weights used in SNNs. This approach allows us to consider device energy consumption during the training phase of SNNs, thereby achieving an energy-aware algorithmic optimization pathway.
\begin{figure*}[htbp]
\centering
\includegraphics[width=0.95\textwidth]{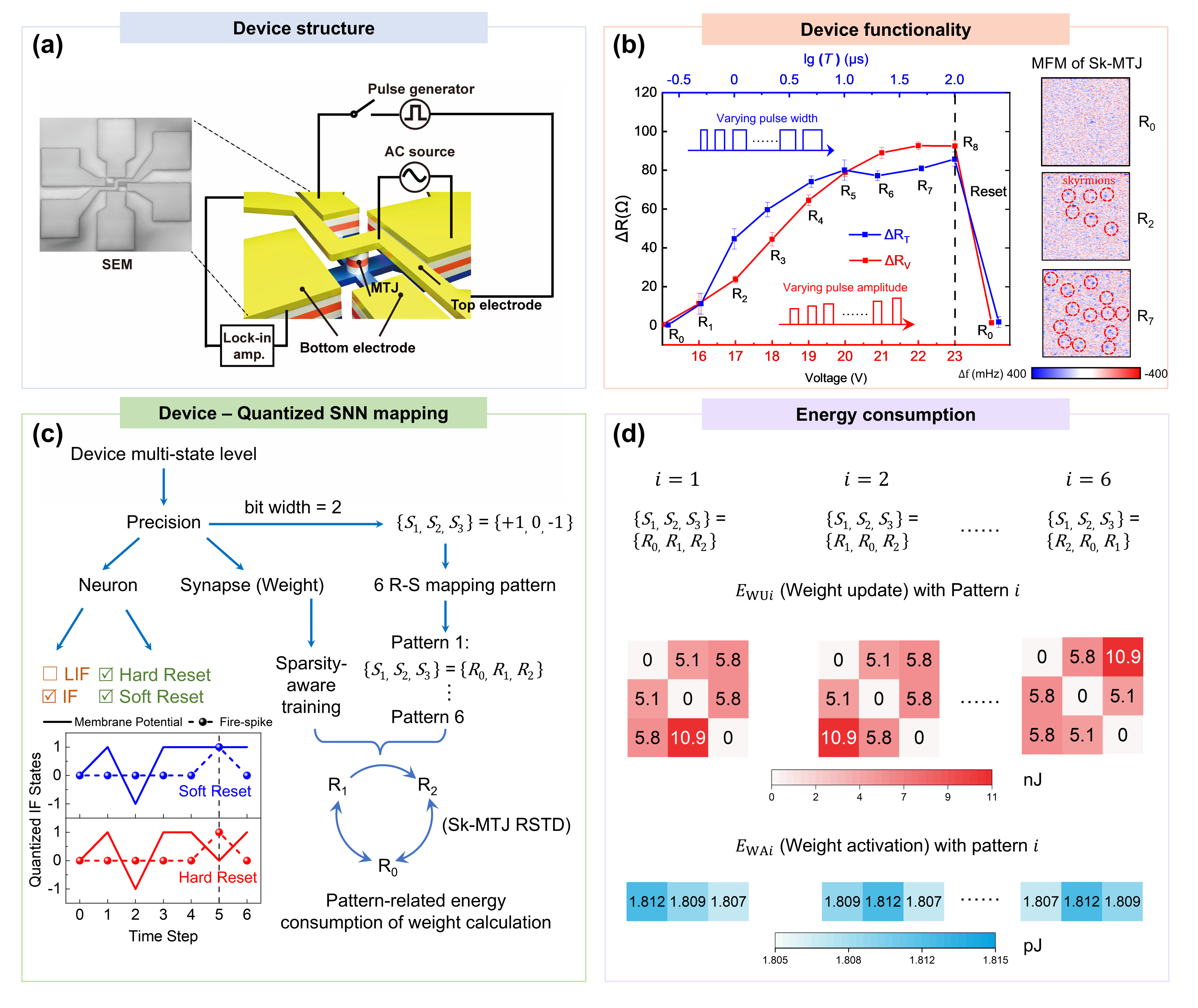}
\caption{(a) Device structure showing the schematic of the Sk-MTJ with measurement setup. (b) Device functionality demonstrated through resistance modulation characteristics under varying pulse conditions, with corresponding MFM images showing the topological skyrmion states in the MTJ causing different resistance states. (c) Device-to-quantized SNN mapping scheme, illustrating the Sk-MTJ capability for neurons and synapses, along with the resistance-state mapping patterns for RSTD implementation. (d) Detailed energy consumption values of one single Sk-MTJ device for both weight update (backpropagation) and weight activation (feedforward) operations across different mapping patterns.}
\label{Fig3}
\end{figure*}

To illustrate our approach, we selected a multi-state spintronic device, specifically the Sk-MTJ, as an example. The resistance state changes in Sk-MTJ are based on the number of skyrmions, which are topologically stable spin structures. This inherent stability helps in better controlling device-to-device variation, making Sk-MTJ an ideal candidate for demonstrating our method.

As shown in Fig.\ref{Fig3}, the design philosophy of the Sk-MTJ neuromorphic device is performed in four steps. 

\textit{Step 1:} Construction of the neuromorphic device. Initially, we fabricate the Sk-MTJ device and establish the electrical measurement setup (Fig. 3a). The Sk-MTJ is designed to be CMOS-compatible, operational at room temperature, and capable of short-pulse resistance state modulation. The scheme of the experimental setup includes Tunnel Magnetoresistance (TMR) measurement and electrical pulse injection for resistance state control. Additionally, a scanning electron microscope (SEM) image of the MTJ device with top and bottom electrodes is provided to give an overview of the device structure.

\textit{Step 2:} Device functionality establishment. The Sk-MTJ resistance can be tuned in two ways: by varying either the pulse width or amplitude (Fig. 3b). Through multiple tests, we have successfully achieved stable tuning of seven discrete resistance states in a stepwise manner. Furthermore, with the assistance of an external magnetic field, the device can be reset from any resistance state to the initial state $R_0$. In practical chip circuit design, pulse amplitude modulation is simpler and faster compared to pulse width modulation. Therefore, we selected the amplitude modulation approach for implementing neuromorphic functionality. Magnetic Force Microscopy (MFM) measurements clearly indicate that the stable resistance changes are directly attributed to the variation in skyrmion density within the device. These tests are essential for validating its applicability in neuromorphic circuits, specifically focusing on the repeatability and stability of resistance changes, which are critical for neuromorphic learning processes.

\textit{Step 3:} Device resistance mapping in Quantized SNN. As shown in Fig. 3c, this step begins by determining the quantization precision based on the seven resistance states obtained in Step 2, which allows for both 2-bit and 3-bit precision. We then assess the neuromorphic functionality achievable at each precision level, using 2-bit precision as an example. Since the device lacks automatic resistance decay, it is well-suited for implementing an IF neuron model, supporting both hard and soft resets. The key distinction between these resets lies in the post-firing membrane potential—soft reset results in a potential less than the threshold, whereas hard reset will definitely bring it to 0. For the synapse model, the mapping pattern must be calculated based on the achievable weight states S. In this case, we identify six possible R-S mapping patterns. Upon examining the actual resistance characteristics of the Sk-MTJ, we observe that only $R_2$. cannot directly transition to $R_1$. This enables us to construct a Resistance State Transition Diagram (RSTD), which serves as the foundation for calculating pattern-related energy consumption in the step 4.

\textit{Step 4:} Energy consumption matrix calculation. For the six R-S mapping patterns, we calculate the energy consumption matrices for both weight update (related to write operations of Sk-MTJ during backpropagation) and weight activation (related to read operations of Sk-MTJ during feedforward propagation), denoted as $E_{\textit{WUi}}$ and $E_{\textit{WAi}}$ for $i=1, 2, ..., 6$. As shown in Fig. 3d, this step is essential for evaluating the energy efficiency of the neuromorphic device and providing a basis for optimizing the QUEST's energy-aware algorithmic design.

\section{Input Encoding}
The human brain, a remarkably complex and efficient spike processing system, employs diverse encoding schemes to represent sensory data. Similarly, in neuromorphic systems, encoding mechanisms are tightly coupled to task types and hardware characteristics. Here, we introduce several encoding schemes used in SNNs, ultimately justifying the choice of direct coding for Sk-MTJ based systems.

(1) \textbf{Rate coding} is widely used, converting input intensity into a spike sequence's mean firing rate. For instance, the Poisson distribution, is a common probabilistic method. While intuitive, rate coding often requires numerous time steps ($> $100) for high accuracy, making it less suitable for real-time online learning. Furthermore, with Sk-MTJ devices, the increased spiking activity of rate coding would lead to excessive energy consumption due to the pulse-controlled nature of their resistances.

(2)\textbf{Temporal (or Latency) Coding} encodes information in the precise timing of single spikes. Although potentially efficient in spike count and effective at capturing temporal patterns, it demands complex decoding mechanisms and precise spike timing. This poses significant challenges for circuit implementation, particularly for Sk-MTJ devices which require stringent control over peripheral clock sources to maintain temporal synchronization and high parallelism. The need for a common time reference across all neurons also raises concerns about signal aliasing. Moreover, effective training algorithms for temporal coding in SNNs remain an open challenge.

(3) \textbf{Direct coding} offers a more efficient alternative by directly mapping input signal intensities to spike trains\cite{wu2019direct}. The first spiking convolutional layer transforms the analog input into a spike train with adaptable time steps. This input encoding layer is co-trained with the entire network, maintaining high accuracy and making it suitable for online learning. For Sk-MTJ devices, direct coding is particularly advantageous, as it allows quantized input signals to precisely modulate the resistance of Sk-MTJs at each time step, enabling accurate control and contributing to significant energy reduction.

As illustrated in Fig. 4, note the input data undergoes repetition across $T$ time steps to accumulate sufficient membrane potential thereby the SNN could perceive inherent image features through temporal dynamics. $T = 4$ is
carefully evaluated to offer a practical and efficient parameters for encoding input signals in SNNs, particularly when dealing with complex datasets for SNN like CIFAR-10 (see visualization of encoding processing in Supplementary Information Fig. S1). In summary, an appropriate direct coding mechanism integration with Sk-MTJ devices enhances the overall system performance, making it a valuable optimization approach in neuromorphic chip.

\begin{figure}[htbp]
\includegraphics[width=1\columnwidth]{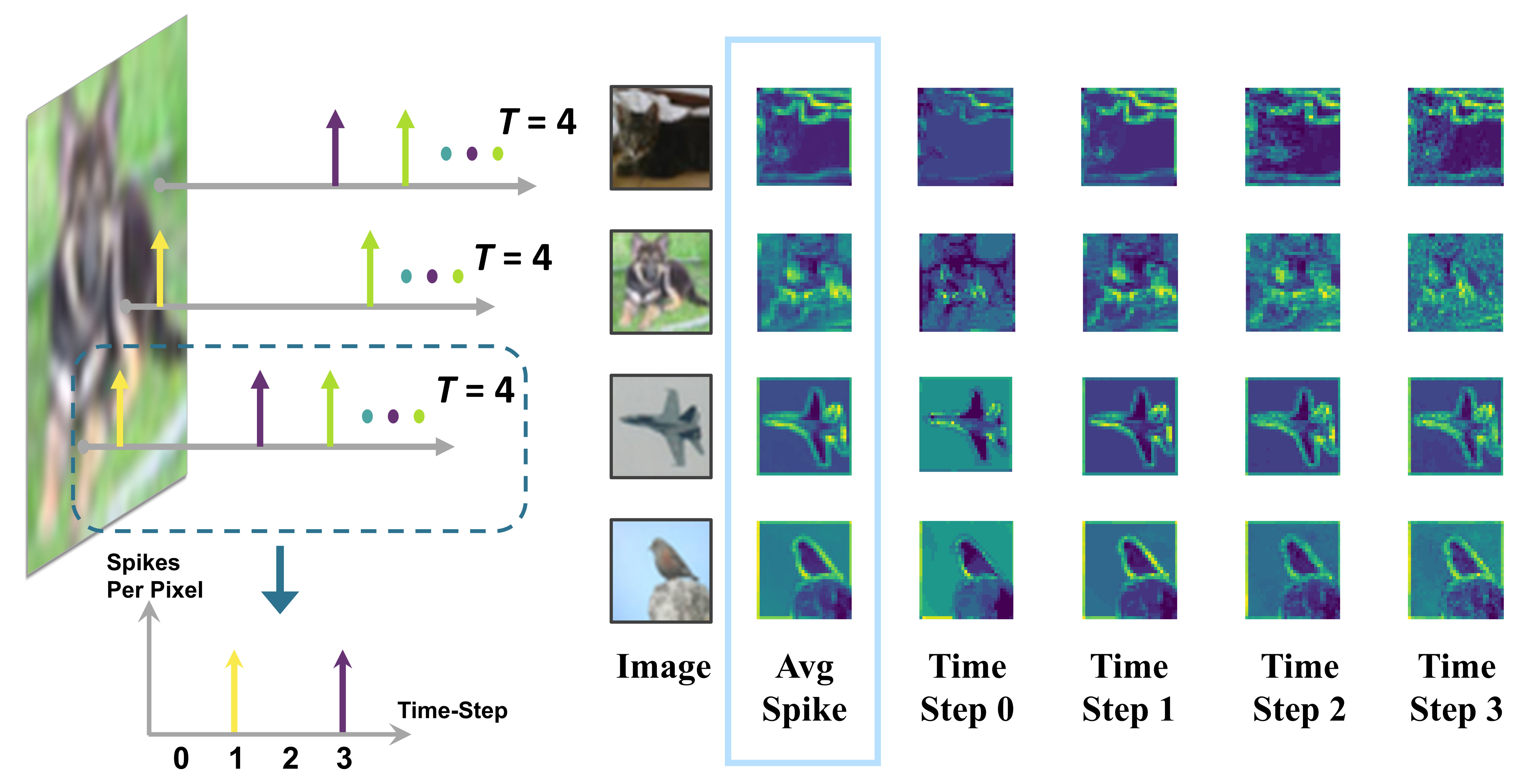}
\caption{Diagram of the direct coding method to encode images. Transformation of image pixels into spike trains using across adaptive T (here T = 4) time steps}
\label{Fig4}
\end{figure}

\section{Quantized Network Training}

Due to the limited resistance states of the Sk-MTJ, representing full precision is not feasible. Therefore, we employ SNN quantization techniques to adapt the network to our device constraints. This approach enables us to reduce the network size while simultaneously reducing energy consumption and minimizing hardware area overhead. Quantization in SNNs has been thoroughly explored in previous studies, which can be classified into three types based on the SNNs’ training scheme. 

(1) \textbf{BPTT-based SNNs} implement two main approaches: optimizing pretrained full-precision networks to reduce weight precision using different methods, such as ADMM\cite{deng2021comprehensive} or K-means clustering\cite{chowdhury2021spatio}; fully integer-based training with fixed-point quantized weights, membrane potentials, and gradients\cite{xie2024toward}.

{(2) \textbf{STDP-based SNNs} are predominantly implemented in shallow networks with local STDP learning\cite{markram1997regulation,mozafari2018first}, which lacks scalability for complex classification tasks.

{(3) \textbf{ANNs-to-SNNs }focus on compressing activations in the original ANNs to enhance energy efficiency and accuracy in SNNs, typically maintaining full or integer precision for weights and membrane potentials\cite{cao2015spiking,rueckauer2017conversion,li2022quantization}.

Since our Sk-MTJ can emulate both synaptic and neuronal functions, it is essential to choose a quantization method that efficiently supports these capabilities. Multiplier-less Integer (MINT)\cite{yin2024mint}, a quantization-aware training method within the BPTT-based SNNs framework, offers simultaneous quantization of both membrane potentials and weights. This approach significantly reduces memory usage for membrane potentials at low quantization bit levels, thus decreasing the overall memory footprint and energy consumption. It is worth noting that, by using a unified scaling factor for both weights and membrane potentials, MINT eliminates the need for multipliers in hardware during inference. This makes it particularly well-suited for on-chip inference in resource-constrained environments.

The QAT implementation of the MINT embedded in QUEST is based on the VGG framework, comprising feedforward and backward propagation phases. As illustrated in Fig. 5, during the forward propagation, input data $X_{in}$ are processed through VGG blocks, which consist of one direct coding block and $L-1$ quantization blocks. The encoding block converts images (FP32) into spikes, which are then fed into the quantization blocks. These quantization blocks are comprised of convolution layers (Conv), followed by neurons, pooling layers (Pool), and fully connected layers (FC) to generate output. Notably, except for the direct coding blocks, the weight data (FP32) will be quantized to $n$ bits through a global quantization (GQ) operation, and the pooling layers are optional components that may not appear in every Spike VGG block.

The backward propagation maintains a similar structure but computes gradients in reverse order, incorporating an inverse quantization operation {GQ}\textsuperscript{-1} to using a straight-through estimator to handle the non-differentiable quantization operations. The training iterates through $L$ VGG blocks ($L$=6 here). The network weights are then updated using these gradients while maintaining the quantization constraints. The key features of our framework include:

(1) A hybrid-precision computation strategy that balances accuracy and efficiency.

(2) An integration contains quantization modules and surrogate-gradient-based backpropagation.

(3) A modular design (adaptive VGG block numbers) enabling framework scalability.

We evaluate our framework on the CIFAR10 dataset, comprising 60,000 32×32 RGB images (50,000 for training, 10,000 for testing). Data augmentation includes random horizontal flipping and clipping. The network is trained using the Adam optimizer with an initial learning rate of $1 \times 10^{-4}$, implementing a step-wise learning rate schedule that decays by a factor of 10 at $50\%$ and $75\%$ of the total epochs (see detailed weight distribution before and after quantization in Supplementary Information Fig. S2). Our experimental results, detailed in the following sections, demonstrate the framework's superior performance while maintaining computational efficiency.

\begin{figure*}[htbp]
\centering
\includegraphics[width=0.9\textwidth, keepaspectratio]{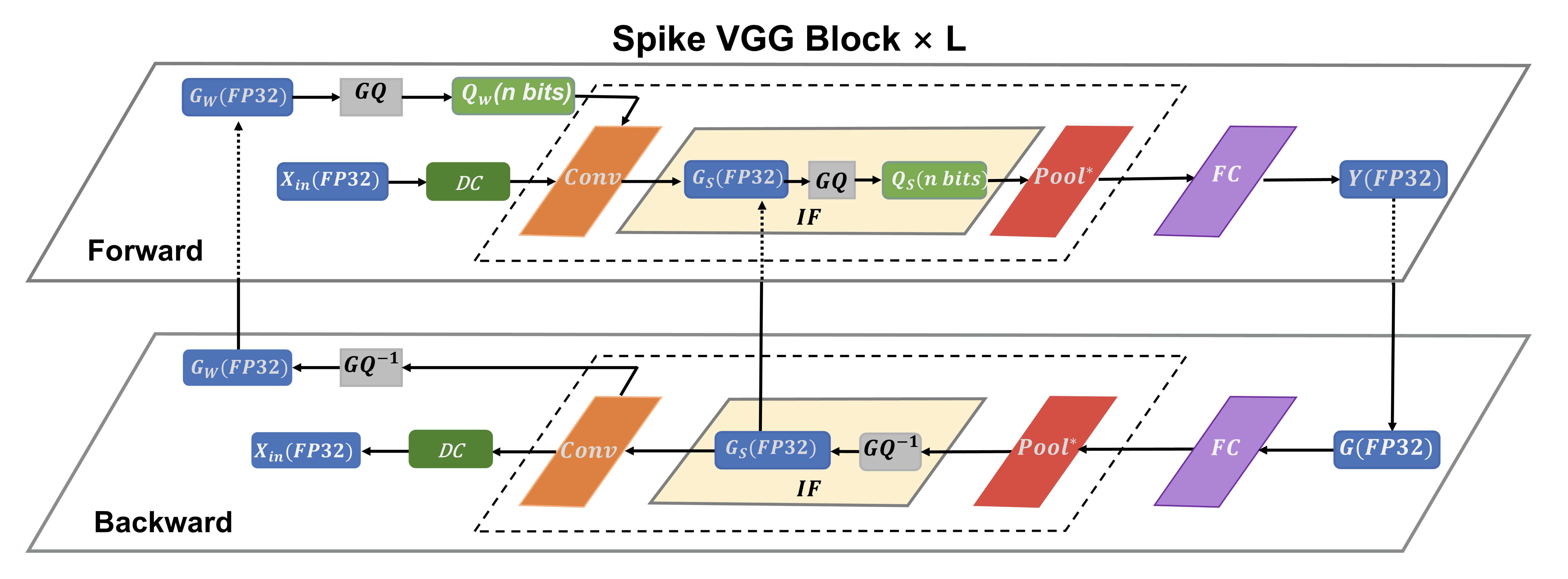}
\caption{Illustration of the quantization architecture showing both forward and backward propagation paths. The forward path processes from input ($X_{in}$) through direct coding (\textit{DC}), convolution (\textit{Conv}), quantization (\textit{GQ}), pooling (\textit{Pool}), and fully connected (\textit{FC}) layers to output. The backward path follows a similar structure with inverse operations for gradient computation. Dashed boxes represent VGG blocks that can be cascaded for deeper architectures.}
\label{Fig5}
\end{figure*}

\section{Results and Discussion}
\begin{figure*}[htbp]
\centering
\includegraphics[width=1\textwidth]{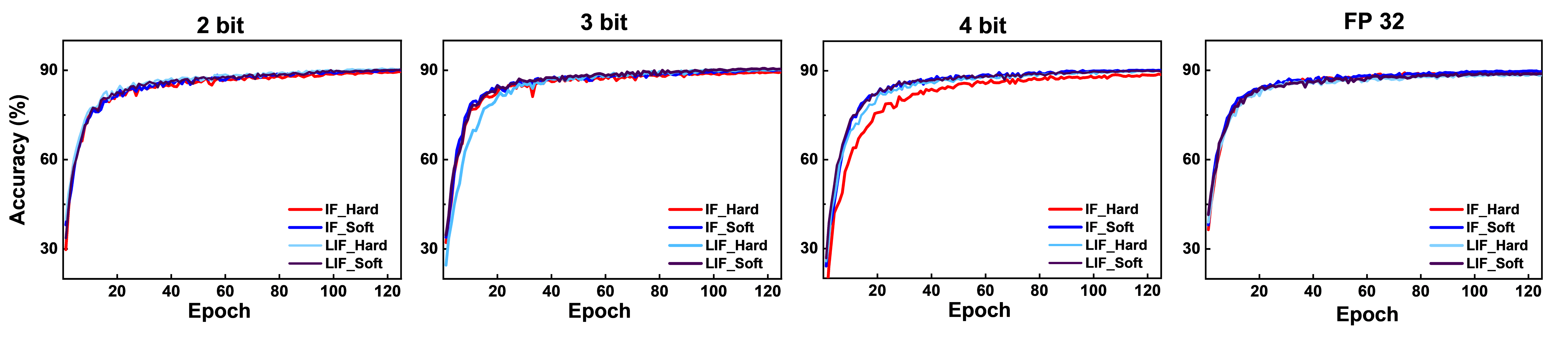}
\caption{Performance comparison of different neuron models ($\mathrm{{\textit{IF\_Hard}}}$, $\mathrm{{\textit{IF\_Soft}}}$, $\mathrm{{\textit{LIF\_Hard}}}$, $\mathrm{{\textit{LIF\_Soft}}}$) across various quantization precisions. The accuracy curves demonstrate the model convergence over training epochs for 2-bit, 3-bit, 4-bit, and full-precision (FP32) implementations.}
\label{Fig6}
\end{figure*}

\subsection{Optimization of Network Configuration}

Firstly, we evaluate the quantization performance of our QUEST training framework across different precisions and neuron models to determine how to co-design the precision and device function. We specifically focus on the accuracy of models employing 2-bit, 3-bit, and 4-bit precision, where both weights and neurons are quantized to the same width, and compare these results to the FP32 baseline. Our findings reveal that the lower precision models maintain competitive accuracy, with a drop of less than $1.5\%$. Moreover, all four neuron models—$\mathrm{{\textit{IF\_Hard}}}$, $\mathrm{{\textit{IF\_Soft}}}$, $\mathrm{{\textit{LIF\_Hard}}}$, and $\mathrm{{\textit{LIF\_Soft}}}$—consistently achieve high accuracy, irrespective of the underlying membrane potential mechanisms. To meet energy efficiency requirements, we adopt a 2-bit precision training scheme along with a hard reset mechanism for neuronal dynamics with an accuracy of 89.6\%. Given the hardware limitations of Sk-MTJ devices, which are incompatible with LIF models, we utilize the IF neuron model instead. This configuration achieves an optimal trade-off between energy efficiency and device practicality while ensuring satisfactory SNN performance.

\subsection{Device-aware Weight Update}

\begin{figure*}[htbp]
\centering
\includegraphics[width=0.85\textwidth]{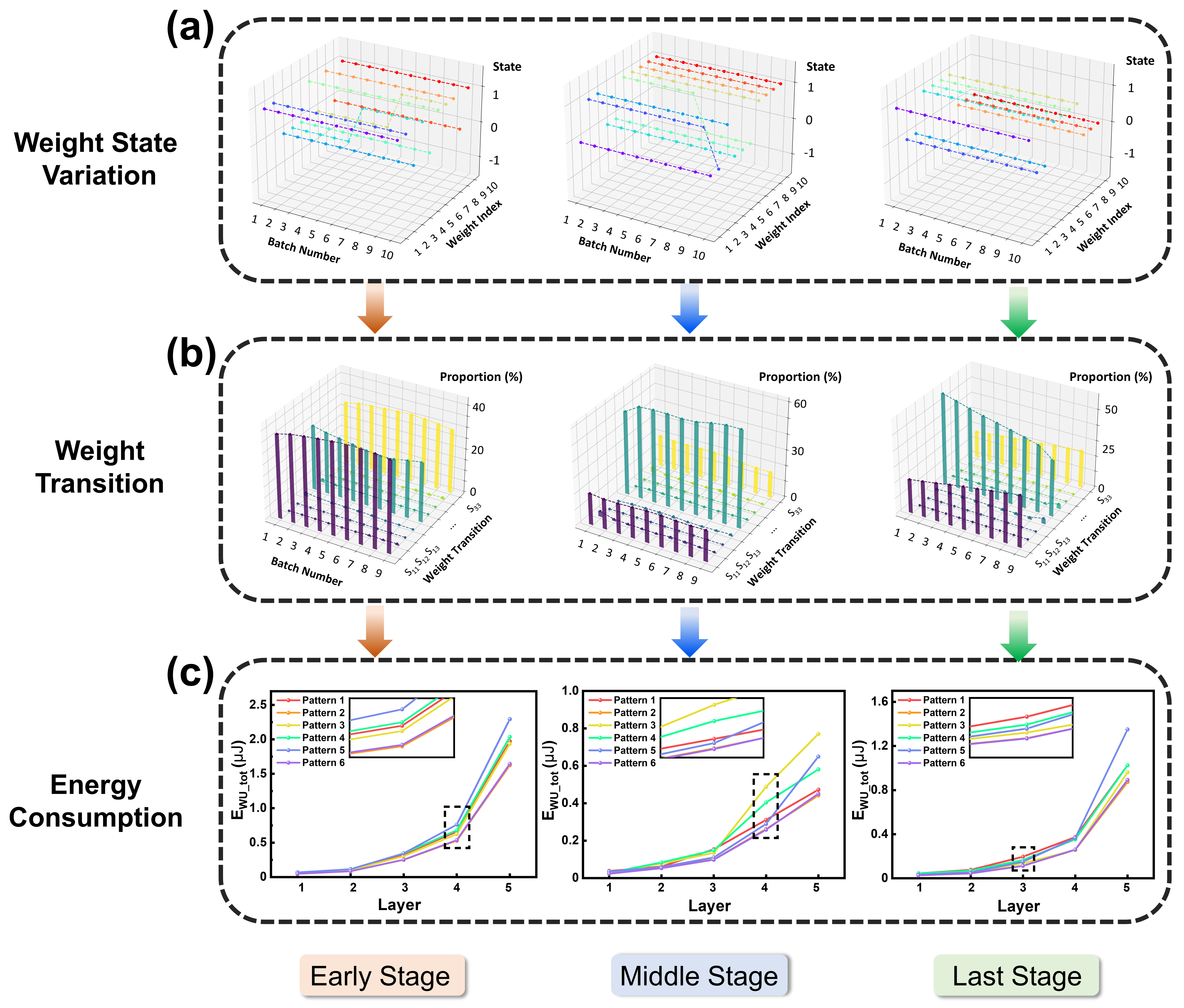}
\caption{Comprehensive analysis of network dynamics across the early, middle, and last training stages. (a) Weight state distribution shows the evolution of quantized 2-bit weights over different batches. (b) Weight transition $S_{\textit{ij}}$ proportions over different batches. (c) Layer-wise weight update energy consumption for different input patterns.}
\label{Fig7}
\end{figure*}
	
Following the establishment of the network architecture, a key step is to optimize the R-S mapping pattern for hardware implementation. To systematically evaluate this optimization, we analyze the network's behavior across three representative stages of the training process. From the total 195 training batches(Batchsize=256), we specifically examine the early stage (batches 1-10), middle stage (batches 96-105), and final stage (batches 180-190).  We firstly introduce an analytical approach that monitors the weight transition process during training, to evaluate layer-wise energy consumption corresponding to different R-S patterns.
	
Fig. 7a shows the evolution of weight states across three stages within the initial epoch, using 10 hierarchically selected weights in the second VGG block (after the first encoding block) as representatives. It reveals that the majority of weights maintain their frozen state throughout the training process. The histograms in Fig. 7b reveal distinct transition patterns, with varying proportions of ($S_{\textit{ij}}$ denotes transition from state $i$ to $j$), by calculating the weights with a monitoring mechanism after each batch input. This mechanism utilizes: (1) a state matrix to store weight states for each batch and (2) a transition matrix to record state changes between batches. Consequently, we obtain the distribution of $S_{\textit{ij}}$ (detailed ratios are available in Supplementary Information Fig. S3). Crucially, even during the initial training stage, where weight adjustments are most frequent, the majority remain unchanged between consecutive batches. This stability is particularly advantageous for NVM implementations, as it allows for the preservation of frozen weights without additional energy overhead. Notably, transitions between non-adjacent states ($S_{\textit{13}}$ or $S_{\textit{31}}$) are absent (0\%), confirming that weight transition occurs exclusively between adjacent values during training.

Fig. 7c presents the layer-wise energy consumption per sample $E_{\textit{WU\_tot}}$ results for six R-S mapping patterns using the $E_{\textit{WU}}$ matrix in Fig. 2d, of which the algorithm is presented in Algorithm 1. To balance computational efficiency and accuracy, we implemented a selective sampling strategy that captures network states from three stages of the first epoch, effectively representing the network's behavior during the whole weight update process. Among all patterns examined, Pattern 6 consistently achieved the lowest \(E_{\textit{WU\_tot}}\) across all layers under three stages, with values generally $\textbf{below 1.65 $\boldsymbol{\mu}$J}$. This is due to the unique property of Pattern 6, where its transition pathway in RSTD coupled with the specific state transition (\(S_{ij}\)) ratios, allows for an optimal energy consumption profile. The observation underscores the significant impact of the R-S mapping pattern on the overall energy efficiency of the neuromorphic system, particularly as network depth increases. It demonstrates that device-aware weight update strategies enable minimizing energy consumption during on-device training of SNNs.
	
	\begin{algorithm}
		\caption{SNN Weight Update Energy Consumption}
        \begin{algorithmic}[1]
        		\Statex \textbf{Input:}
        		\Statex \hspace{0.5cm} $E_{\textit{WU}}[ij]$: Energy consumption matrix for weight update
        		\Statex \hspace{0.5cm} $T^e_{b,l}[ij]$: Weight transition matrix for epoch $e$, batch $b$, layer $l$
                \Statex \hspace{0.5cm} $i, j$: Matrix element row and column indices
        		\Statex \hspace{0.5cm} $N_l$: Number of weights in layer $l$.
        		\Statex \hspace{0.5cm} $E$: Number of epochs.
        		\Statex \hspace{0.5cm} $B$: Number of batches.
        		\Statex \hspace{0.5cm} $L$: Number of layers.
        		\Statex \hspace{0.5cm} Batchsize: Number of samples in a batch.
        		\Statex \textbf{Output:}
        		\Statex \hspace{0.5cm} $\text{Energy}^e_{l}$: Energy consumption matrix
        		\For{each epoch $e \in [1, \ldots, E]$}
        		\For{each batch $b \in [1, \ldots, B]$}
        		\For{each layer $l \in [1, \ldots, L]$}
        		\State Initialize $T^e_{b,l}[ij] \gets 0$
        		\State $T^e_{b,l} \gets$ Compute transition matrix for current epoch, batch, and layer
        		\State $T^e_{b,l}[ij] \gets T^e_{b,l}[ij] / N_l$ 
        		\EndFor
        		\EndFor
        		\For{$l=1$ to $L$}
        		\State $\text{Energy}^e_{l}[ij]$= $\sum_{b=1}^{B} (T^e_{b,l}[ij] * E_{\textit{WU}}[ij])$ / Batchsize
        		\EndFor
        		\EndFor
        		\State \textbf{Return} $\text{Energy}^e_{l}$ for all $e$ and $l$
        	\end{algorithmic}
	\end{algorithm}

	\begin{figure*}[t]
		\centering
		\includegraphics[width=1\textwidth]{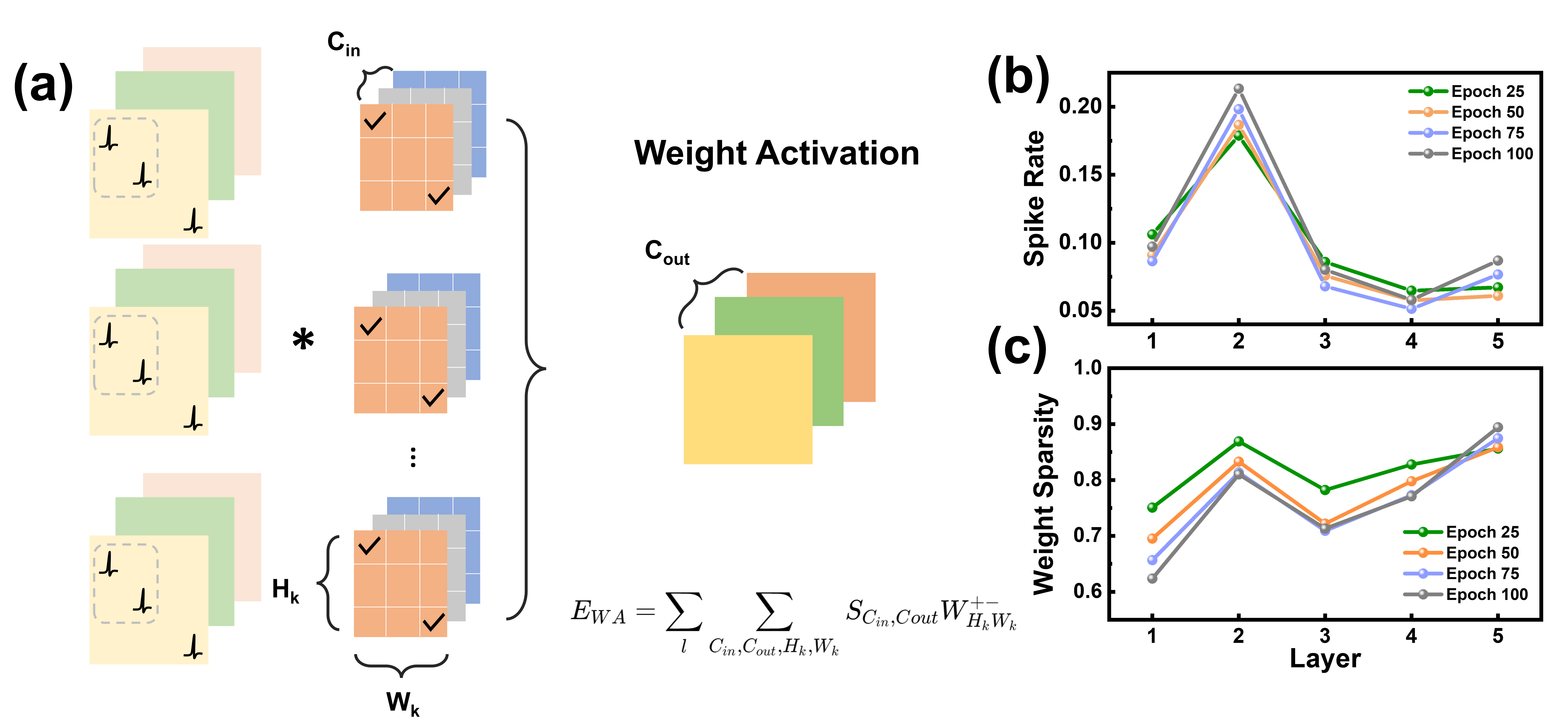}
		\caption{The illustration of spike rate and weight sparsity used in the calculated weight activation. (a) The weight activation with the convolution process with input features ($C_{in}$), kernel weights height($H_k$) and width($W_k$), and output features ($C_{out}$). (b) Layer-wise spike rate across different training epochs. A spike rate of {n} indicates that every neuron fired on average {n} times for each sample over all time steps (here time steps is 4). (c) Layer-wise weight sparsity across different training epochs. A weight sparsity of {m} signifies that {m} percent of the zero-value weights.}
		\label{Fig8}
	\end{figure*}
    
\subsection{Energy-Efficient Weight Activation}
	In this section, we evaluate the energy consumption of weight activation, which only considers the non-zero weight activation when spikes (value = 1) occur, cutting down on NVM read operations during the feedforward process without sacrificing accuracy, as shown in Fig. 8a. In this process, the inputs are convolved with corresponding $H_k\times W_k$ kernels, generating a single-channel feature map and then aggregating to multiple feature maps. Fig. 8b depicts the spike rate (also known as the firing rate) across different epochs, as specified by 
	\begin{align}
		S^l = \frac{S^l_{\text{total}}}{N^l}, \quad \text{where } S^l_{\text{total}} = \sum_{t=1}^{4} S^l_{t} \label{eq:sl}
	\end{align}
	where $S^l$ is the average spike rate of the $l$-th layer, $S^l_{\text{total}}$ is the total number of spikes and $N^l$ is the number of neurons in the $l$-th layer. $S^l=1$ indicates that the SNN performs the same number of operations as an equivalent ANN. Averaging across layers 1 to 5, QUEST achieved a remarkably low average spike rate of 0.13 at epoch 100 during training. This demonstrates significant inherent sparsity, leading to fewer spike operations of the Sk-MTJ neural device served as neurons.

Furthermore, our investigation reveals that energy efficiency in SNNs extends beyond mere spike rate considerations. Operations involving zero-valued weights are deemed computationally ineffective as they do not contribute to the network's output, highlighting the critical role of weight sparsity in quantized SNNs. This dual relationship manifests through reduced memory footprint, decreased access frequency, and simplified hardware implementation. Fig. 8c illustrates the evolution of weight sparsity across the layers and training epochs, $W^l$ is defined as the proportion of zero-valued weights within the $l$-th layer:
	\begin{align}
		W^l = \frac{\sum_{i=1}^{N^l} \delta(w_i)}{N^l}, \text{where } \delta(w_i) = \begin{cases} 1, w_i = 0 \\0,w_i = \pm 1\end{cases} \label{eq:wl}
	\end{align}
	$N^l$ denotes the total number of weights in the $l$-th layer. $\delta(w_i)$ is an indicator function that determines whether the weight $w_i$ layer is 0. Our results demonstrate that QUEST quantization significantly impacts weight sparsity, with over 60\% of weights becoming zero-valued across layers as epochs increase. Notably, the $5$-th layer exhibits the highest weight sparsity $\sim$85\%. This high weight sparsity, combined with a low spike rate, contributes to the high energy efficiency observed in weight activation – the process of reading operations of the synaptic Sk-MTJs served as weights. To quantify this efficiency, we introduce, for the first time, the \textbf{Activation Operation Rate (AOR)} metric, defined as:
	\begin{align}
		{\text{AOR}}^l = \frac{\text{OP}^l_{\textit{eff}}}{\text{OP}^l_{\textit{tot}}} \label{eq:aosl}
	\end{align}
${\text{AOR}}^l$ represents the proportion of effective operations (where both a spike is present AND the weight is non-zero) out of the total operations in the $l$-th layer. ${\text{OP}}^l_{\text{eff}}$ and ${\text{OP}}^l_{\text{tot}}$ denote the number of effective and total operations, respectively. Fig. 9a reveals the results of very low AOR ($< 0.081$) in different layers and epochs, especially in the deeper $2$-$5$ layers. It yields a $\sim$\,4.7$\times$ rate decrease over the spike rate.

Subsequently, the per-sample energy consumption for weight activation, denoted as $E_{\textit{WA\_tot}}$, is calculated using the $E_{\textit{WA}}$ of Pattern 6 (Fig. 2d), as detailed in Algorithm 2. Calculations for other patterns are provided in Supplementary information Fig. S4. Notably, $E_{\textit{WA\_tot}}$ exhibits a trend closely resembling that of AOR across layers and epochs, as illustrated in Fig. 9b.  The first layer exhibits the highest value in the range of $\textbf{54-89 $\boldsymbol{\mu}$J}$, and the $5$-th layer show the lowest value in the range of $\textbf{3-5 $\boldsymbol{\mu}$J}$. 

To evaluate the energy efficiency of QUEST, we compared its activation energy cost using operation rate with that of the ANNs and SNNs in 45 nm CMOS technology, adopting the same calculation method in \cite{rathi2021diet} (see Supplementary information Table 1). We found that QUEST achieves approximately $\sim$93$\times$, $\sim$2.47$\times$ and $\sim$7.1$\times$ energy efficiency improvement over ANN, VGG6-DIET-SNN and VGG16-DIET-SNN, respectively. This remarkable energy efficiency enhancement can be attributed to QUEST's sparsely quantized SNN architecture, which leverages the inherently low AOR and minimal activation energy $E_{\textit{WA}}$ of the Sk-MTJ device.

	\begin{figure*}[t]
		\centering
		\includegraphics[width=1\textwidth]{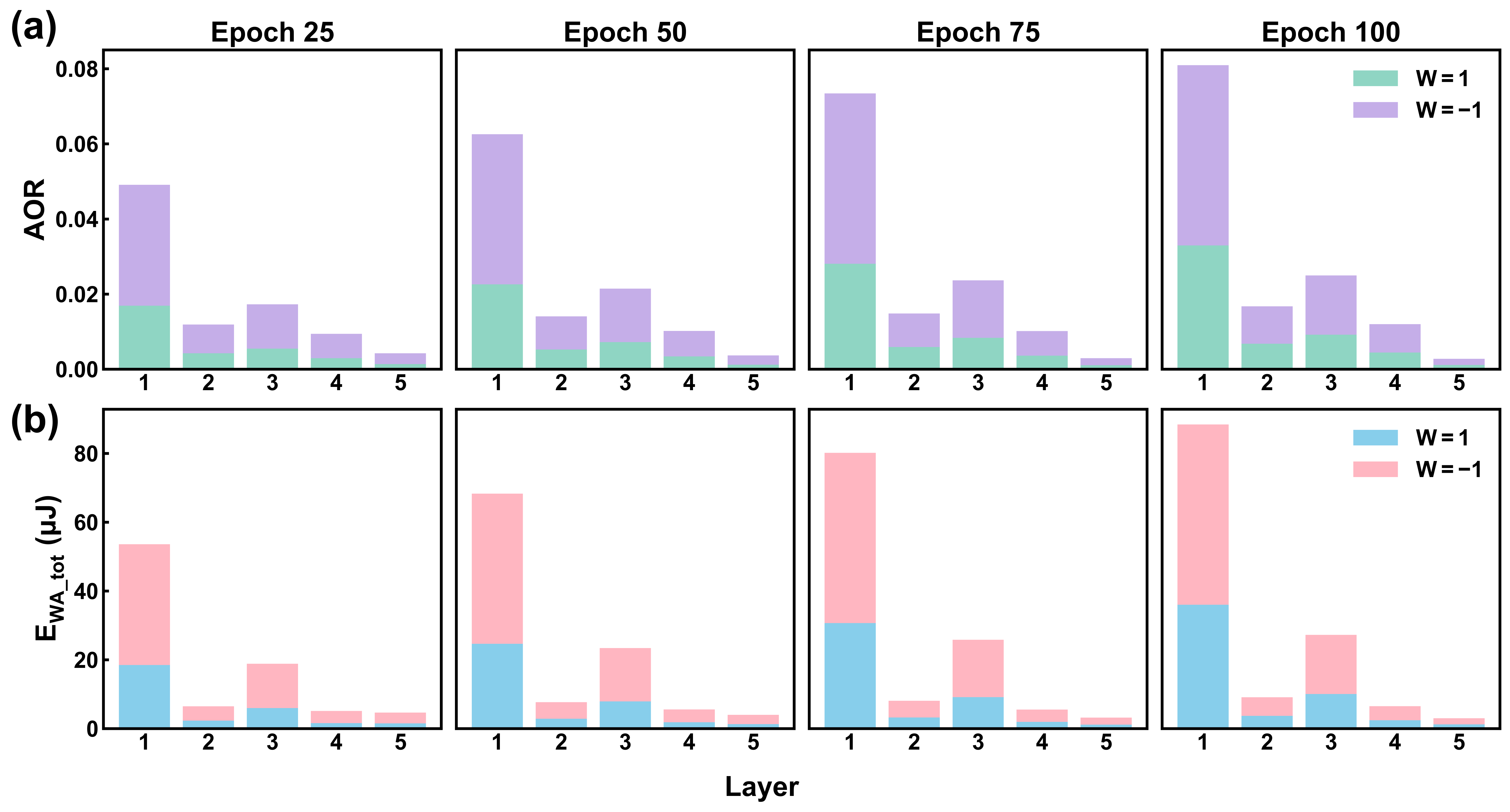}
		\caption{Results of AOR and weight activation energy consumption using pattern 6. (a) Layer-wise AOR during training across different training epochs. (b) Layer-wise weight activation energy consumption across different training epochs. Both include different non-zero weight states ($W = -1$ and $W = 1$)}
		\label{Fig9}
	\end{figure*}

	\section{Conclusion}
	In this article, we introduce QUEST, a novel co-design framework for energy-aware SNN training on multi-state neuromorphic devices. A key feature of QUEST is device-aware optimization tailored specifically for NVM devices, which successfully achieves trade-offs among high accuracy, low bit-precision and scalable SNNs. Notably, through our newly proposed R-S mapping pattern and AOR metric, QUEST demonstrates significant energy efficiency improvements in NVM-based neuromorphic systems. While this work focuses on specific cases within each module (multi-state NVM device, encoding mechanism, quantization, neuron models, and network topologies), it can be further expanded and adapted to fit diverse on-device training and algorithmic demands.
	
	By bridging the gap between device-level constraints and algorithmic requirements through synergistic optimization methodologies, QUEST not only offers practical design guidelines for scalable SNN implementation, but also facilitates new perspectives for developing cross-layer interfaces in neuromorphic chips.
	
	\begin{algorithm}[t]
		\caption{SNN Weight Activation Energy Consumption}
		\begin{algorithmic}[1]
		\Statex \textbf{Input:}
		\Statex \hspace{0.5cm} $W^t_l[ij]$: Quantized weight of layer $l$ at time step $t$
		\Statex \hspace{0.5cm} $S^t_{l-1}[ij]$: Input spike of layer $l-1$ at time step $t$
		\Statex \hspace{0.5cm} Weight Activation energy $E^+$, $E^-$ $\in$ $M_{\textbf{WA}}$
		\Statex \hspace{0.5cm} $T$: Number of time steps.
		\Statex \hspace{0.5cm} $L$: Number of layers.
		\Statex \hspace{0.5cm} $i, j$: Matrix element row and column indices
		\Statex \textbf{Output:}
		\Statex \hspace{0.5cm} $\text{count}^t_{l,+}$: Number of multiplications where weight is +1 and input spike.
		\Statex \hspace{0.5cm} $\text{count}^t_{l,-}$: Number of multiplications where weight is -1 and input spike.
		\Statex \hspace{0.5cm} $\text{count}^t_{l,\textit{ineff}}$: Number of other multiplications.
		\Statex \hspace{0.5cm} $\text{OP}^l_{\textit{eff}}$: Number of effective multiplications for layer $l$.
		\Statex \hspace{0.5cm} $\text{OP}^l_{\textit{tot}}$: Number of total multiplications for layer $l$.
		\For{$l = 1$ to $L$}
		\For{$t = 1$ to $T$}
		\State Initialize $\text{count}^t_{l,+} \gets 0$
		\State Initialize $\text{count}^t_{l,-} \gets 0$
		\State Initialize $\text{count}^t_{l,\textit{ineff}} \gets 0$
		\For{each element $[i, j]$ in $S^t_{l-1}$ and $W^t_l$}
		\If{$S^t_{l-1}[ij] = 1$}
		\If{$W^t_l[ij] = +1$}
		\State $\text{count}^t_{l,+} \gets \text{count}^t_{l,+} + 1$
		\ElsIf{$W^t_l[ij] = -1$}
		\State $\text{count}^t_{l,-} \gets \text{count}^t_{l,-} + 1$
		\Else
		\State $\text{count}^t_{l,\textit{ineff}} \gets \text{count}^t_{l,\textit{ineff}} + 1$
		\EndIf
		\EndIf
		\EndFor
		\EndFor
		\EndFor
		\Statex \textbf{Return} \{
		\begin{align*}
			&\text{OP}^l_{\textit{eff}} = (\sum_{t=1}^{T}\text{count}^t_{l,+} + \sum_{t=1}^{T}\text{count}^t_{l,-}), \\
			&\text{OP}^l_{\textit{tot}} = (\sum_{t=1}^{T}\text{count}^t_{l,+} + \sum_{t=1}^{T}\text{count}^t_{l,-} + \sum_{t=1}^{T}\text{count}^t_{l,\textit{ineff}}), \\
			&\text{AOR} = \frac{\text{OP}^l_{\textit{eff}}}{\text{OP}^l_{\textit{tot}}}, \\
			&E_{\textit{WA\_tot}} = \frac{\sum_{t=1}^{T}\text{count}^t_{l,+}}{\text{OP}^l_{\textit{tot}}}*E^+ + \frac{\sum_{t=1}^{T}\text{count}^t_{l,-}}{\text{OP}^l_{\textit{tot}}}*E^-,
		\end{align*} \} for all $l$
	\end{algorithmic}
	\end{algorithm}
	
	\bibliographystyle{IEEEtran}
	\bibliography{ref}

\begin{thebibliography}{10}
\providecommand{\url}[1]{#1}
\csname url@samestyle\endcsname
\providecommand{\newblock}{\relax}
\providecommand{\bibinfo}[2]{#2}
\providecommand{\BIBentrySTDinterwordspacing}{\spaceskip=0pt\relax}
\providecommand{\BIBentryALTinterwordstretchfactor}{4}
\providecommand{\BIBentryALTinterwordspacing}{\spaceskip=\fontdimen2\font plus
\BIBentryALTinterwordstretchfactor\fontdimen3\font minus \fontdimen4\font\relax}
\providecommand{\BIBforeignlanguage}[2]{{%
\expandafter\ifx\csname l@#1\endcsname\relax
\typeout{** WARNING: IEEEtran.bst: No hyphenation pattern has been}%
\typeout{** loaded for the language `#1'. Using the pattern for}%
\typeout{** the default language instead.}%
\else
\language=\csname l@#1\endcsname
\fi
#2}}
\providecommand{\BIBdecl}{\relax}
\BIBdecl

\bibitem{eshraghian2023training}
J.~K. Eshraghian, M.~Ward, E.~O. Neftci, X.~Wang, G.~Lenz, G.~Dwivedi, M.~Bennamoun, D.~S. Jeong, and W.~D. Lu, ``Training spiking neural networks using lessons from deep learning,'' \emph{Proceedings of the IEEE}, 2023.

\bibitem{hassabis2017neuroscience}
D.~Hassabis, D.~Kumaran, C.~Summerfield, and M.~Botvinick, ``Neuroscience-inspired artificial intelligence,'' \emph{Neuron}, vol.~95, no.~2, pp. 245--258, 2017.

\bibitem{maass1997networks}
W.~Maass, ``Networks of spiking neurons: the third generation of neural network models,'' \emph{Neural networks}, vol.~10, no.~9, pp. 1659--1671, 1997.

\bibitem{frenkel2023bottom}
C.~Frenkel, D.~Bol, and G.~Indiveri, ``Bottom-up and top-down approaches for the design of neuromorphic processing systems: tradeoffs and synergies between natural and artificial intelligence,'' \emph{Proceedings of the IEEE}, vol. 111, no.~6, pp. 623--652, 2023.

\bibitem{rathi2021diet}
N.~Rathi and K.~Roy, ``Diet-snn: A low-latency spiking neural network with direct input encoding and leakage and threshold optimization,'' \emph{IEEE Transactions on Neural Networks and Learning Systems}, vol.~34, no.~6, pp. 3174--3182, 2021.

\bibitem{zhang2021tuning}
T.~Zhang, S.~Jia, X.~Cheng, and B.~Xu, ``Tuning convolutional spiking neural network with biologically plausible reward propagation,'' \emph{IEEE Transactions on Neural Networks and Learning Systems}, vol.~33, no.~12, pp. 7621--7631, 2021.

\bibitem{zhang2020fully}
X.~Zhang, Z.~Wu, J.~Lu, J.~Wei, J.~Lu, J.~Zhu, J.~Qiu, R.~Wang, K.~Lou, Y.~Wang \emph{et~al.}, ``Fully memristive snns with temporal coding for fast and low-power edge computing,'' in \emph{2020 IEEE International Electron Devices Meeting (IEDM)}.\hskip 1em plus 0.5em minus 0.4em\relax IEEE, 2020, pp. 29--6.

\bibitem{kim2023c}
S.~Kim, S.~Kim, S.~Hong, S.~Kim, D.~Han, and H.-J. Yoo, ``C-dnn: A 24.5-85.8 tops/w complementary-deep-neural-network processor with heterogeneous cnn/snn core architecture and forward-gradient-based sparsity generation,'' in \emph{2023 IEEE International Solid-State Circuits Conference (ISSCC)}.\hskip 1em plus 0.5em minus 0.4em\relax IEEE, 2023, pp. 334--336.

\bibitem{pei2019towards}
J.~Pei, L.~Deng, S.~Song, M.~Zhao, Y.~Zhang, S.~Wu, G.~Wang, Z.~Zou, Z.~Wu, W.~He \emph{et~al.}, ``Towards artificial general intelligence with hybrid tianjic chip architecture,'' \emph{Nature}, vol. 572, no. 7767, pp. 106--111, 2019.

\bibitem{davies2018loihi}
M.~Davies, N.~Srinivasa, T.-H. Lin, G.~Chinya, Y.~Cao, S.~H. Choday, G.~Dimou, P.~Joshi, N.~Imam, S.~Jain \emph{et~al.}, ``Loihi: A neuromorphic manycore processor with on-chip learning,'' \emph{Ieee Micro}, vol.~38, no.~1, pp. 82--99, 2018.

\bibitem{merolla2014million}
P.~A. Merolla, J.~V. Arthur, R.~Alvarez-Icaza, A.~S. Cassidy, J.~Sawada, F.~Akopyan, B.~L. Jackson, N.~Imam, C.~Guo, Y.~Nakamura \emph{et~al.}, ``A million spiking-neuron integrated circuit with a scalable communication network and interface,'' \emph{Science}, vol. 345, no. 6197, pp. 668--673, 2014.

\bibitem{grollier2020neuromorphic}
J.~Grollier, D.~Querlioz, K.~Camsari, K.~Everschor-Sitte, S.~Fukami, and M.~D. Stiles, ``Neuromorphic spintronics,'' \emph{Nature electronics}, vol.~3, no.~7, pp. 360--370, 2020.

\bibitem{zhang2018sign}
Q.~Zhang, H.~Wu, P.~Yao, W.~Zhang, B.~Gao, N.~Deng, and H.~Qian, ``Sign backpropagation: An on-chip learning algorithm for analog rram neuromorphic computing systems,'' \emph{Neural Networks}, vol. 108, pp. 217--223, 2018.

\bibitem{lele2023heterogeneous}
A.~S. Lele, M.~Chang, S.~D. Spetalnick, B.~Crafton, S.~Konno, Z.~Wan, A.~Bhat, W.-S. Khwa, Y.-D. Chih, M.-F. Chang \emph{et~al.}, ``A heterogeneous rram in-memory and sram near-memory soc for fused frame and event-based target identification and tracking,'' \emph{IEEE Journal of Solid-State Circuits}, 2023.

\bibitem{yang2024fully}
Z.~Yang, W.~Yue, C.~Liu, Y.~Tao, P.~J. Tiw, L.~Yan, Y.~Yang, T.~Zhang, B.~Dang, K.~Liu \emph{et~al.}, ``Fully hardware memristive neuromorphic computing enabled by the integration of trainable dendritic neurons and high-density rram chip,'' \emph{Advanced Functional Materials}, p. 2405618, 2024.

\bibitem{wang2021phase}
Q.~Wang, G.~Niu, W.~Ren, R.~Wang, X.~Chen, X.~Li, Z.-G. Ye, Y.-H. Xie, S.~Song, and Z.~Song, ``Phase change random access memory for neuro-inspired computing,'' \emph{Advanced Electronic Materials}, vol.~7, no.~6, p. 2001241, 2021.

\bibitem{jiao2020monatomic}
F.~Jiao, B.~Chen, K.~Ding, K.~Li, L.~Wang, X.~Zeng, and F.~Rao, ``Monatomic 2d phase-change memory for precise neuromorphic computing,'' \emph{Applied Materials Today}, vol.~20, p. 100641, 2020.

\bibitem{tang2018ecram}
J.~Tang, D.~Bishop, S.~Kim, M.~Copel, T.~Gokmen, T.~Todorov, S.~Shin, K.-T. Lee, P.~Solomon, K.~Chan \emph{et~al.}, ``Ecram as scalable synaptic cell for high-speed, low-power neuromorphic computing,'' in \emph{2018 IEEE International Electron Devices Meeting (IEDM)}.\hskip 1em plus 0.5em minus 0.4em\relax IEEE, 2018, pp. 13--1.

\bibitem{luo2019capacitor}
J.~Luo, L.~Yu, T.~Liu, M.~Yang, Z.~Fu, Z.~Liang, L.~Chen, C.~Chen, S.~Liu, S.~Wu \emph{et~al.}, ``Capacitor-less stochastic leaky-fefet neuron of both excitatory and inhibitory connections for snn with reduced hardware cost,'' in \emph{2019 IEEE International Electron Devices Meeting (IEDM)}.\hskip 1em plus 0.5em minus 0.4em\relax IEEE, 2019, pp. 6--4.

\bibitem{yan2023moire}
X.~Yan, Z.~Zheng, V.~K. Sangwan, J.~H. Qian, X.~Wang, S.~E. Liu, K.~Watanabe, T.~Taniguchi, S.-Y. Xu, P.~Jarillo-Herrero \emph{et~al.}, ``Moir{\'e} synaptic transistor with room-temperature neuromorphic functionality,'' \emph{Nature}, vol. 624, no. 7992, pp. 551--556, 2023.

\bibitem{zhang2023low}
A.~Zhang, J.~Shi, J.~Wu, Y.~Zhou, and W.~Yu, ``Low latency and sparse computing spiking neural networks with self-driven adaptive threshold plasticity,'' \emph{IEEE Transactions on Neural Networks and Learning Systems}, 2023.

\bibitem{burkitt2006review}
A.~N. Burkitt, ``A review of the integrate-and-fire neuron model: I. homogeneous synaptic input,'' \emph{Biological cybernetics}, vol.~95, pp. 1--19, 2006.

\bibitem{massarotto2024adiabatic}
M.~Massarotto, S.~Saggini, M.~Loghi, and D.~Esseni, ``Adiabatic leaky integrate and fire neurons with refractory period for ultra low energy neuromorphic computing,'' \emph{npj Unconventional Computing}, vol.~1, no.~1, pp. 1--11, 2024.

\bibitem{yao2023tunneling}
Y.~Yao, H.~Cheng, B.~Zhang, J.~Yin, D.~Zhu, W.~Cai, S.~Li, and W.~Zhao, ``Tunneling magnetoresistance materials and devices for neuromorphic computing,'' \emph{Materials Futures}, vol.~2, no.~3, p. 032302, 2023.

\bibitem{wang2020neuromorphic}
S.~Wang, X.~Chen, X.~Huang, D.~Wei~Zhang, and P.~Zhou, ``Neuromorphic engineering for hardware computational acceleration and biomimetic perception motion integration,'' \emph{Advanced Intelligent Systems}, vol.~2, no.~11, p. 2000124, 2020.

\bibitem{bhattacharya2023reram}
T.~Bhattacharya, S.~Bezugam, S.~Pande, E.~Wlazlak, and D.~Strukov, ``Reram-based neohebbian synapses for faster training-time-to-accuracy neuromorphic hardware,'' in \emph{2023 International Electron Devices Meeting (IEDM)}.\hskip 1em plus 0.5em minus 0.4em\relax IEEE, 2023, pp. 1--4.

\bibitem{peng2020dnn+}
X.~Peng, S.~Huang, H.~Jiang, A.~Lu, and S.~Yu, ``Dnn+ neurosim v2. 0: An end-to-end benchmarking framework for compute-in-memory accelerators for on-chip training,'' \emph{IEEE Transactions on Computer-Aided Design of Integrated Circuits and Systems}, vol.~40, no.~11, pp. 2306--2319, 2020.

\bibitem{chen2017neurosim+}
P.-Y. Chen, X.~Peng, and S.~Yu, ``Neurosim+: An integrated device-to-algorithm framework for benchmarking synaptic devices and array architectures,'' in \emph{2017 IEEE International Electron Devices Meeting (IEDM)}.\hskip 1em plus 0.5em minus 0.4em\relax IEEE, 2017, pp. 6--1.

\bibitem{pehle2021norse}
C.-G. Pehle and J.~Egholm~Pedersen, ``Norse-a deep learning library for spiking neural networks,'' \emph{Zenodo}, 2021.

\bibitem{fang2023spikingjelly}
W.~Fang, Y.~Chen, J.~Ding, Z.~Yu, T.~Masquelier, D.~Chen, L.~Huang, H.~Zhou, G.~Li, and Y.~Tian, ``Spikingjelly: An open-source machine learning infrastructure platform for spike-based intelligence,'' \emph{Science Advances}, vol.~9, no.~40, p. eadi1480, 2023.

\bibitem{vetter2023abisko}
J.~S. Vetter, P.~Date, F.~Fahim, S.~R. Kulkarni, P.~Maksymovych, A.~A. Talin, M.~G. Tallada, P.~Vanna-Iampikul, A.~R. Young, D.~Brooks \emph{et~al.}, ``Abisko: Deep codesign of an architecture for spiking neural networks using novel neuromorphic materials,'' \emph{The International Journal of High Performance Computing Applications}, vol.~37, no. 3-4, pp. 351--379, 2023.

\bibitem{roy2024spintronic}
K.~Roy, C.~Wang, S.~Roy, A.~Raghunathan, K.~Yang, and A.~Sengupta, ``Spintronic neural systems,'' \emph{Nature Reviews Electrical Engineering}, pp. 1--16, 2024.

\bibitem{hamilton2021best}
T.~Hamilton, ``The best of both worlds,'' \emph{Nature Machine Intelligence}, vol.~3, no.~3, pp. 194--195, 2021.

\bibitem{zhang2024hyb}
F.~Zhang, L.~Yang, and D.~Fan, ``Hyb-learn: A framework for on-device self-supervised continual learning with hybrid rram/sram memory,'' in \emph{Proceedings of the 61st ACM/IEEE Design Automation Conference}, 2024, pp. 1--6.

\bibitem{li2022experimental}
S.~Li, A.~Du, Y.~Wang, X.~Wang, X.~Zhang, H.~Cheng, W.~Cai, S.~Lu, K.~Cao, B.~Pan \emph{et~al.}, ``Experimental demonstration of skyrmionic magnetic tunnel junction at room temperature,'' \emph{Science Bulletin}, vol.~67, no.~7, pp. 691--699, 2022.

\bibitem{wu2019direct}
Y.~Wu, L.~Deng, G.~Li, J.~Zhu, Y.~Xie, and L.~Shi, ``Direct training for spiking neural networks: Faster, larger, better,'' in \emph{Proceedings of the AAAI conference on artificial intelligence}, vol.~33, no.~01, 2019, pp. 1311--1318.

\bibitem{deng2021comprehensive}
L.~Deng, Y.~Wu, Y.~Hu, L.~Liang, G.~Li, X.~Hu, Y.~Ding, P.~Li, and Y.~Xie, ``Comprehensive snn compression using admm optimization and activity regularization,'' \emph{IEEE transactions on neural networks and learning systems}, vol.~34, no.~6, pp. 2791--2805, 2021.

\bibitem{chowdhury2021spatio}
S.~S. Chowdhury, I.~Garg, and K.~Roy, ``Spatio-temporal pruning and quantization for low-latency spiking neural networks,'' in \emph{2021 International Joint Conference on Neural Networks (IJCNN)}.\hskip 1em plus 0.5em minus 0.4em\relax IEEE, 2021, pp. 1--9.

\bibitem{xie2024toward}
H.~Xie, G.~Yang, and W.~Gao, ``Toward efficient deep spiking neuron networks: A survey on compression,'' in \emph{International Joint Conference on Artificial Intelligence}.\hskip 1em plus 0.5em minus 0.4em\relax Springer, 2024, pp. 18--31.

\bibitem{markram1997regulation}
H.~Markram, J.~L{\"u}bke, M.~Frotscher, and B.~Sakmann, ``Regulation of synaptic efficacy by coincidence of postsynaptic aps and epsps,'' \emph{Science}, vol. 275, no. 5297, pp. 213--215, 1997.

\bibitem{mozafari2018first}
M.~Mozafari, S.~R. Kheradpisheh, T.~Masquelier, A.~Nowzari-Dalini, and M.~Ganjtabesh, ``First-spike-based visual categorization using reward-modulated stdp,'' \emph{IEEE transactions on neural networks and learning systems}, vol.~29, no.~12, pp. 6178--6190, 2018.

\bibitem{cao2015spiking}
Y.~Cao, Y.~Chen, and D.~Khosla, ``Spiking deep convolutional neural networks for energy-efficient object recognition,'' \emph{International Journal of Computer Vision}, vol. 113, pp. 54--66, 2015.

\bibitem{rueckauer2017conversion}
B.~Rueckauer, I.-A. Lungu, Y.~Hu, M.~Pfeiffer, and S.-C. Liu, ``Conversion of continuous-valued deep networks to efficient event-driven networks for image classification,'' \emph{Frontiers in neuroscience}, vol.~11, p. 682, 2017.

\bibitem{li2022quantization}
C.~Li, L.~Ma, and S.~Furber, ``Quantization framework for fast spiking neural networks,'' \emph{Frontiers in Neuroscience}, vol.~16, p. 918793, 2022.

\bibitem{yin2024mint}
R.~Yin, Y.~Li, A.~Moitra, and P.~Panda, ``Mint: Multiplier-less integer quantization for energy efficient spiking neural networks,'' in \emph{2024 29th Asia and South Pacific Design Automation Conference (ASP-DAC)}.\hskip 1em plus 0.5em minus 0.4em\relax IEEE, 2024, pp. 830--835.

\end{thebibliography}

\end{document}